\documentclass[usenatbib]{mnras}

\usepackage{newtxtext,newtxmath}
\usepackage{multirow}
\usepackage{bm}
\usepackage{xcolor}

\usepackage[T1]{fontenc}

\DeclareRobustCommand{\VAN}[3]{#2}
\let\VANthebibliography\thebibliography
\def\thebibliography{\DeclareRobustCommand{\VAN}[3]{##3}\VANthebibliography}

\usepackage{graphicx}	
\usepackage{amsmath}	

\title[Transits around polluted white dwarfs]{The frequency of transiting planetary systems around polluted white dwarfs}

\author[A.~Robert et al.]{Akshay Robert,$^{1}$\thanks{E-mail: akshay.robert.22@ucl.ac.uk}
Jay Farihi,$^1$ 
Vincent Van Eylen,$^2$ 
Amornrat Aungwerojwit,$^3$ 
Boris T.~G\"{a}nsicke,$^4$
\newauthor
Seth Redfield,$^5$
Vikram S.~Dhillon,$^{6,7}$ 
Thomas R.~Marsh,$^4$\thanks{Deceased}
and Andrew Swan$^4$ 
\\
$^{1}$Department of Physics and Astronomy, University College London, London WC1E 6BT, UK\\
$^{2}$Mullard Space Science Laboratory, University College London, Dorking RH5 6NT, UK\\
$^{3}$Department of Physics, Naresuan University, Phitsanulok 65000, Thailand\\
$^{4}$Department of Physics, University of Warwick, Coventry CV4 7AL, UK\\
$^{5}$Astronomy Department and Van Vleck Observatory, Wesleyan University, Middletown, CT 06459, USA\\
$^{6}$Department of Physics and Astronomy, University of Sheffield, Sheffield S3 7RH, UK\\
$^{7}$Instituto de Astrof\'isica de Canarias, E-38205 La Laguna, Spain
}

\date{Accepted XXX. Received YYY; in original form ZZZ}

\pubyear{2023}

\begin{document}
\label{firstpage}
\pagerange{\pageref{firstpage}--\pageref{lastpage}}
\maketitle

\begin{abstract}
This paper investigates the frequency of transiting planetary systems around metal-polluted white dwarfs using high-cadence photometry from ULTRACAM and ULTRASPEC on the ground, and space-based observations with {\em TESS}.  Within a sample of 313 metal-polluted white dwarfs with available {\em TESS} light curves, two systems known to have irregular transits are blindly recovered by box-least-squares and Lomb-Scargle analyses, with no new detections, yielding a transit fraction of $0.8_{-0.4}^{+0.6}$~per cent. Planet detection sensitivities are determined using simulated transit injection and recovery for all light curves, producing upper limit occurrences over radii from dwarf to Kronian planets, with periods from 1\,h to 27\,d.  The dearth of short-period, transiting planets orbiting polluted white dwarfs is consistent with engulfment during the giant phases of stellar evolution, and modestly constrains dynamical re-injection of planets to the shortest orbital periods. Based on simple predictions of transit probability, where $(R_* + R_{\rm p})/a\simeq0.01$, the findings here are nominally consistent with a model where 100~per cent of polluted white dwarfs have circumstellar debris near the Roche limit;  however, the small sample size precludes statistical confidence in this result.
Single transits are also ruled out in all light curves using a search for correlated outliers,  providing weak constraints on the role of Oort-like comet clouds in white dwarf pollution.

\end{abstract}



\section{Introduction}

Transit photometry is the leading method of detecting extrasolar planets, and is responsible for several thousand discoveries that constitute the bulk of confirmed detections \citep{lissauer18-1, budrikis22-1}. From a geometric standpoint, the transit method favours the detection of planets on orbits close to their host stars, whereas sensitivity favours larger transit depths, such as those caused by giant and dwarf gas planets.

Relative to main-sequence stars, white dwarfs provide two potential advantages in transit searches: their small radii and the ubiquitous presence of closely orbiting planetary debris. Compared to the sun, while a typical white dwarf radius $R_*=0.013\,R_{\odot}$ decreases the nominal transit probability by two orders of magnitude, it produces a 50~per cent transit depth for an Earth-size planet, and thus for similar photometric precision yields a sensitivity gain of nearly four orders of magnitude. Unlike main-sequence stars, there is comprehensive evidence of planetesimal debris discs orbiting within roughly $1\,R_{\odot}$ of metal-polluted white dwarfs \citep{reachetal05-3, juraetal07-1}. At face value, this predicts around 1~per cent of polluted white dwarfs may exhibit transits. 

At least 25~per cent of white dwarfs exhibit signatures of planetary systems, which include emission from circumstellar dust and gas \citep{farihietal09-1,girvenetal12-1,manseretal20-1}, photospheric metals that require ongoing replenishment \citep{zuckermanetal10-1, koesteretal14-1}, and transits \citep{vanderburgetal15-1,vanderburgetal20-2}.  In support of these observations, there are theoretical predictions for planets closely orbiting white dwarfs, possibly resulting from debris disc evolution \citep{vanlieshout18-1}, or post-main sequence dynamics \citep{veras+gaensicke15-1}.

Transits provide a distinct view of white dwarf planetary systems, with a modest but growing number of discoveries \citep[see][]{guidryetal21-1}. These have revealed new empirical insights, and confirmed inferences based on prior work.  The rings of debris can have semimajor axes significantly larger than a solar radius, and thus potentially be highly eccentric \citep{vanderboschetal20-1}, where some exhibit line of sight gas absorption \citep{xuetal16-1,redfieldetal17-1}.  The debris discs are highly dynamic, with large scale heights \citep{gaensickeetal16-1,izquierdoetal18-1}, and therefore distinct from the canonical picture based on the rings of Saturn \citep{rappaportetal16-1,cauleyetal18-1}. White dwarfs with transits generally do not have infrared excesses, suggesting their detection results from favourable viewing geometry and not enhanced dust production \citep{farihietal22-1}. 

Currently, white dwarf transits fall into two categories: those with regular transits from self-gravitating bodies \citep[WD\,1856+534;][]{vanderburgetal20-2}, and irregular transits consistent with circumstellar debris fields, where metals have been detected when high-resolution spectroscopy is available \citep[e.g.\ ZTF\,J0328--1219, WD\,1054--226, WD\,1145+017;][]{vanderburgetal15-1,vanderboschetal21-1,farihietal22-1}.  While all white dwarfs with infrared dust emission exhibit photospheric metals \citep{farihi16-1}, some that exhibit transits do not, or currently lack decisive data \citep{guidryetal21-1}.

Given that white dwarf pollution requires the ongoing or recent deposition of circumstellar debris, it may be inevitable that some exhibit transits. However, despite key forerunner investigations of transiting planets around white dwarfs \citep{faedietal11-1, fultonetal14-1, vansluijs+vaneylen18-1}, the transit frequency toward polluted white dwarfs is currently unconstrained, and in particular for transiting debris discs. This work aims to constrain the frequency of planetary systems around metal-polluted white dwarfs, and utilizes standard methodology of box-least-squares and simulated injection and recovery for high-cadence photometric observations of 33 targets from the ground, and 313 objects from space.  The observations are described in Section~2, and the transit signal search and sensitivity are detailed in Section~3.  The results are discussed in Section~4, including a determination of transit frequency and comparison to related surveys, with a brief summary and outlook in Section~5.

\section{Observations and Data}

The data used in this paper are based on two programs targeting polluted white dwarfs: the first is a ground-based transit search for a sample of 33 stars using the high-speed photometry cameras ULTRACAM and ULTRASPEC \citep{dhillonetal07-1,dhillonetal14-1}, and the second analyzes {\em Transiting Exoplanet Survey Satellite} ({\em TESS}; \citealt{rickeretal15-1}) {\sc pdcsap} light curves for 313 objects. 
The latter targets are a bright subsample that represent around 20~per cent of all known polluted white dwarfs, and were identified by cross-matching literature sources with the {\em TESS} Input Catalogue (TIC v8.2; \citealt{stassunetal19-1}).

\subsection{ULTRACAM and ULTRASPEC}\label{ULTRACAM_obs}

High-speed photometry for 19 polluted white dwarfs was obtained with ULTRACAM, a triple-beam imaging camera that enables simultaneous acquisition of data in wavelength regions covering the $u$, $g$, and $riz$ bands, and facilitates rapid capture of the field of view via frame transfer technology. The observations were carried out using super-SDSS filters $u_{\rm s}g_{\rm s}r_{\rm s}$ (with higher throughput compared to the conventional filters; \citealt{dhillonetal21-1}), between 2017 Nov 17 and 2021 Aug 20 at La Silla Observatory, where ULTRACAM is mounted on the 3.5-m New Technology Telescope.

These data were reduced using the HiPERCAM pipeline software \citep{dhillonetal21-1}.  Differential photometry was performed on the source and at least one comparison star using a variable aperture (the number of field stars available for normalisation ranged from one to five depending on the field). The frames were corrected for bias and flat-fielded using normalized sky flats obtained in a continuous spiral (to remove stars) during twilight. To maximise the signal-to-noise ratio and the sampling cadence in the final light curves used here for transit detection, only the $g$ and $r$ bands are used in the subsequent analysis. These data were then co-added to further improve the fidelity of the final light curves, with a median scatter in the individual $g$ and $r$ bands of 1.2 and 1.7~per cent, respectively.

A further 14 polluted white dwarfs were observed with ULTRASPEC, which is mounted on the 2.4-m Thai National Telescope at Doi Inthanon. ULTRASPEC operates similarly to ULTRACAM, but instead utilises a single channel.  The data were obtained with a Schott KG5 filter to maximise light throughput, where the transmission approximates the wavelengths spanned by $ugr$ combined. The ULTRASPEC observations were taken between 2015 Nov 30 and 2020 Jan 09, and the data were reduced as for ULTRACAM. The median scatter in ULTRASPEC target star light curves is 2.5~per cent. The time stamps for light curves from both instruments were converted to Barycentric Julian Day using Barycentric Dynamical Time.  All ground-based target observations are summarized in Table~\ref{tab:obs}.

\subsection{\textit{TESS} Data}\label{TESS_obs}

{\em TESS} has surveyed the entire sky for transiting planets since its launch in 2018. During science operations, its four cameras take nearly continuous exposures, from which full-frame images and target pixel files of selected targets are produced by a dedicated data pipeline. All available light curves from Sectors 1--65 with 120\,s-cadence data were downloaded via the Science Processing Operations Center pipeline. In addition, 214 of these stars had 20\,s-cadence light curves, which were downloaded and analyzed in the same fashion as the 120\,s data. This sample had mean {\em TESS} light curve baselines of 2.34 and 1.72 sectors for the 120 and 20\,s cadences, respectively.  When downloading the light curves, only data with pipeline {\tt quality} flags of 0 or 512 (impulsive outlier flag) were accepted. Despite typically indicating anomalous data, the impulsive outlier flag was accepted, based on the short durations expected for white dwarf transits (less than 2\,min for a transiting Earth analogue, and thus only one or two in-transit measurements for the 120\,s-cadence data). 

The source photometry obtained from \textit{TESS} is susceptible to crowding owing to the 21\,arcsec size and response function of the camera pixels. For exoplanet transits, there will be a reduction in the observed depths in blended light curves, leading to underestimated planet radii. This problem is worse for white dwarfs as they are intrinsically faint, where the \textit{TESS} estimation of crowding, encoded by the \textsc{crowdsap} parameter, ranges from 0.011 to 0.998 (median and standard deviation $0.53\pm0.32$) for stars in this study.  Despite the correction of any flux dilution in the {\sc pdcsap} light curves \citep{jenkinsetal16-1}, crowding often results in false positive detections of variability and transit-like signals (see Section~\ref{transit_search}).

\begin{figure}
\includegraphics[width=\columnwidth]{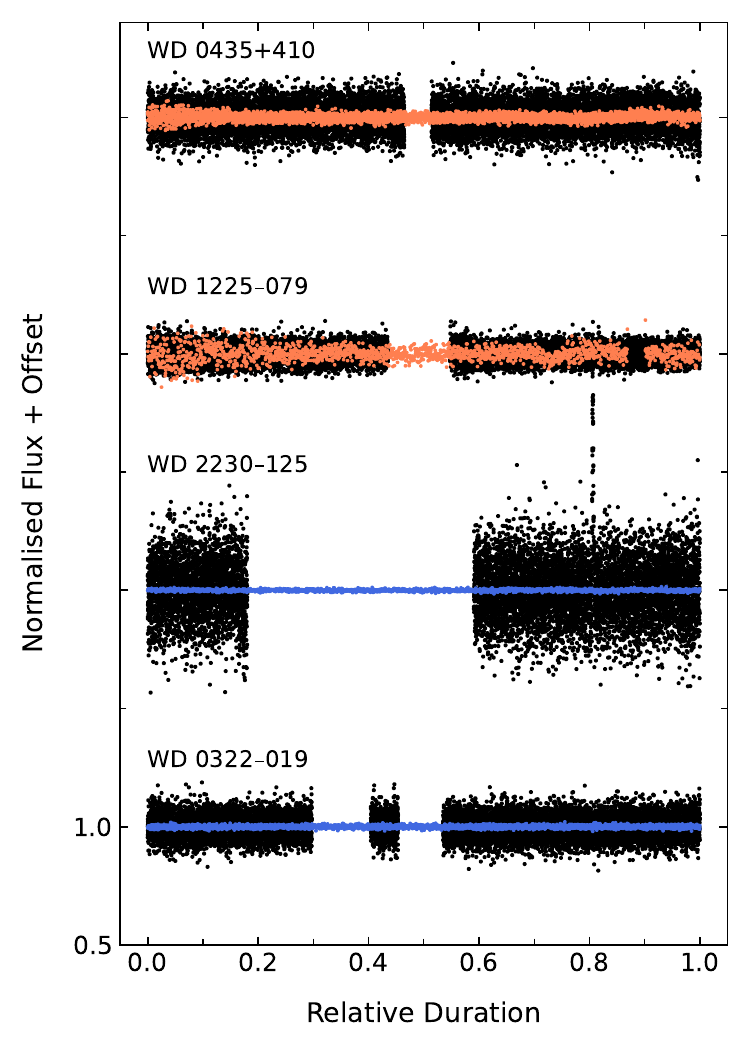}
\caption{A representative selection of two ULTRACAM (blue) and two ULTRASPEC (orange) light curves, each plotted over one sector of their {\em TESS} 120\,s data (black). The $x$-axis has been adjusted to normalize the full duration of each light curve. The spike in the {\em TESS} light curve of WD\,2230--125 is caused by a passing asteroid, resulting in temporary brightening. These examples highlight the difference in photometric scatter between the ULTRACAM (median 0.5~per cent), ULTRASPEC  (median 2.5~per cent), and {\em TESS} (median 12.5~per cent) light curves.}
\label{ucam_uspec_TESS_LCs}
\end{figure}

All long-term trends were removed from the light curves using a Tukey bi-weight slider from the \textsc{w\={o}tan} package \citep{hippkeetal19-1}, where a kernel size of 0.2\,d was used for most of the data; this is approximately a factor of three larger than the maximum expected transit duration for a planet with radius 2\,$R_{\rm Jup}$ and period 27\,d.  In special cases where the light curve displayed rapid variability (owing to contamination from nearby sources), a smaller, tailored kernel size was adopted.  The de-trended {\em TESS} light curves have significantly larger scatter than for the ground-based observations, with a median of 12.5~per cent, illustrated in Fig.~\ref{ucam_uspec_TESS_LCs}.

\section{Methods and Data Analysis}\label{s3}

The following section details the algorithms used to search for new transiting systems, and the procedures 
to determine (upper limit) occurrences for planetary systems, in the case of either real or null detections. First, a transit search was executed for all light curves using the box-least-squares algorithm \citep[BLS;][]{kovacsetal02-1}, which identifies exoplanet transit signals by fitting boxcar functions to light curves phase folded on trial periods. Second, transit injection and recovery was used to determine the sensitivities of the respective instruments to planetary transit signals.

\subsection{The transit search algorithm}\label{transit_search}

For the ULTRACAM and ULTRASPEC light curves, where the median light curve duration is 4.1\,h, BLS was used to search for transit signals in a period range spanning 15\,min to 7\,h, with a grid spacing of 3.6\,s, where the upper period bound is slightly less than twice the median, ground-based observation baseline.  The simulated period range for the ground-based data encompasses orbits from the inner edge of a debris disc to just outside the Roche limit. For the {\em TESS} sample, periods from 1\,h to 27\,d, spaced by 60\,s, were searched for both the 120 and 20\,s cadence light curves. The lower bound of 1\,h is interior to the Roche limit for a white dwarf, and the upper bound of 27\,d corresponds to the approximate baseline of a single {\em TESS} sector. In both cases, the period grid is sufficiently fine to detect short-duration transits, while also enabling calculation of the BLS periodogram for each light curve at reasonable computational expense. From the periodogram, three candidate periods that pass the detection threshold are selected. The first candidate period corresponds to the signal with the highest power in the BLS spectrum, and likewise the second and third candidate periods correspond to the next two respective maxima in the periodogram, with the condition that they are neither harmonics of the best candidate period nor of each other. 


Owing to the irregular dimming events in the light curves of white dwarfs with transiting debris, the Lomb-Scargle algorithm was used in addition to BLS, to facilitate the detection of any possible periodic signals. Both algorithms were used to perform a blind search in both the ground-based and \textit{TESS} samples. For all light curves, the Lomb-Scargle periodogram was computed for the raw data, and BLS was run after de-trending. 

The detection thresholds for all light curves are individually determined using a bootstrap randomisation technique \citep[e.g.][]{greissetal14-1,hermesetal15-1}. In short, for a given light curve, the fluxes are randomly shuffled while preserving the original time stamps of the observations. The periodogram is subsequently re-computed for each randomised light curve to extract the signal with the highest peak in the power spectrum. This randomisation process is repeated a total of 1000 times per light curve, from which the 99.9$^{\rm th}$ percentile power is calculated, and this value is adopted as the detection threshold for that light curve.  The median signal detection efficiencies (defined in eq. 6 of \citealt{kovacsetal02-1}) and standard deviations of the detection thresholds are $10.8\pm2.9$ for the {\em TESS} light curves, and $12.1\pm3.7$ for the ground-based data. 

All {\em TESS} targets were inspected for contamination owing to aperture crowding using {\sc tess-localize} \citep{higgins+bell23-1}, which determines the source of variability in the sky for a given target pixel file and set of frequencies. The input frequencies were informed by the output of the Lomb-Scargle periodogram for each light curve. All periodograms and phase-folded light curves for each candidate period were visually examined for each target, with particular attention given to stars that were determined to be variable by {\sc tess-localize}. Visual inspection may introduce uncertainties in the characterisation of light curve variability, but is unlikely to have a significant effect on the overall results. There were 32 instances where signals in the Lomb-Scargle periodogram passed the associated false alarm threshold, but that were subsequently determined to be spurious by {\sc tess-localize}. This corresponds to a 9.6~per cent contaminant rate for variability detections in the space-based sample.

\subsection{Transit injection and recovery}

In an ideal scenario, the planetary occurrence is simply the ratio of the number of detected systems to the total sample size $N$. However, only a proportion $Np_{\rm transit} = N \cdot (R_* + R_{\rm p})/a$ of planetary systems transit the stellar disc, and within this reduced subset, only a fraction $S(R_{\rm p}, P)$ -- the detection sensitivity -- can be found with a given instrument. Both of these factors hinder the detection of transiting systems in a photometric survey, and in the case of a null detection, yield only upper limits on their actual frequency. To account for both of these, an effective sample size {\em N'} can be used instead \citep{faedietal11-1}, defined as 
\begin{equation}
    N'(R_{\rm p}, P) = N \cdot p_{\rm transit}(R_{\rm p}, P) \cdot S(R_{\rm p}, P) \: ,
\end{equation}
\noindent
and this approach is used to determine all transit occurrences hereafter. 

The detection sensitivity encodes the probability that an instrument can detect planets of a given period and radius, and can be computed via the method of injection and recovery. This process injects a simulated transit signal into a light curve, for which a BLS periodogram is computed using the injected data. Because the period and radius of the synthetic transit are known quantities, if the correct period or a harmonic is identified by BLS, then a real signal (i.e.\ regular transits from a self-gravitating body or irregular dimming events from debris) of the same period and radius should have also been detected. The sensitivity is then simply the ratio of the number of recovered injections to the total number of simulations. The Python software package \textsc{batman} \citep{kreidberg15-1} was used to inject transit signals into the de-trended light curves where no pre-existing transit features were present (32 ground-based, 311 and 213 {\em TESS} 120\,s- and 20\,s-cadence light curves, respectively). Because transits were injected following light curve de-trending, it may be possible that the resulting detection sensitivities are marginally higher than otherwise. All sample stars were assumed to have $R_*=0.013\,R_{\odot}$ and $M_*=0.65$\,M$_{\odot}$, with uniform limb darkening, and the impact parameter of the synthetic planet was drawn from a uniform distribution spanning 0 to $(R_{\rm p} + R_*)/R_*$ (i.e.\ grazing transits are permitted). The injected transits had zero eccentricity: on the one hand, tides are theorised to circularise orbits within a few Myr for planets that survive until the white dwarf phase \citep{nordhaus+spiegel13-1}; on the other hand, the data provide no actual constraints on eccentricity in the absence of detections. 
The model integration time was chosen to match the exposure times for the ground-based observations and {\em TESS} cadences, respectively. 

To initialise the injection and recovery process, a period-radius grid was defined using eight logarithmically-spaced radii from 0.0625 to 8.00 ${\rm R}_\oplus$ (corresponding to dwarf planets like Ceres up to the radius of Saturn). The injected period range for each sample was identical to the period grid used in the respective transit searches in Section~\ref{transit_search}. For each injection, a light curve was randomly selected from the sample, and values for the period and radius of the synthetic transit signal were randomly drawn from a uniform distribution. A synthetic transit signal was considered to be recovered by the search algorithm if the best-fit and injected periods were within a tolerance of 2~per cent. The simulations were run until the recovered sensitivity converged to within 0.5~per cent of the local median value within each period-radius cell, with a total of 37\,800 injections performed for each of the ground-based and {\em TESS} 120\,s cadence light curves, and 20\,400 injections for the {\em TESS} 20\,s subset (fewer in the latter case, to reduce computational expense).

\subsection{Occurrence constraints}

If transits are detected, the occurrence in radius-period phase space is the number of detections $n$ divided by the effective sample size $N'$ \citep{howardetal12-1, dressing+charbonneau13-1}. In the case that no transits are found, upper limit occurrences can be calculated as a function of radius and period. Assuming a binomial distribution describes the probability of finding $n$ planetary systems in an effective sample of $N'$ white dwarfs \citep{burgasseretal03-2, faedietal11-1}, the rate of occurrence for a null detection is given by \citep{vansluijs+vaneylen18-1}

\begin{equation}\label{max_occurence_null}
f_{\rm max} = 1 - (1-C)^{\frac{1}{1+N'}} ,
\end{equation}

where $f_{\rm max}$ is the maximum occurrence for a null detection at an arbitrary confidence level $C$. 
The uncertainty in effective sample size is 
\begin{equation}\label{sigma_effective_sample}
    \sigma_{N'} = \sqrt{(p_{\rm transit} \times N \times \sigma_{S})^2 + (S \times N \times \sigma_{\rm transit})^2} ,
\end{equation}

where $\sigma_{S}$ and $\sigma_{\rm transit}$ are the uncertainties on the detection sensitivity and transit probability, respectively. The uncertainty on the maximum occurrence $f_{\rm max}$ for a null detection is therefore
\begin{equation}\label{sigma_max_occurrence_null}
    \sigma_{f_{\rm max}} = \left| \frac{\partial f_{\rm max}}{\partial N'} \, \sigma_{N'} \right| = \left| \frac{ln(1-C) \cdot (1-C)^{1+N'}}{(1+N')^2} \, \sigma_{N'} \right| .
\end{equation}
As a compromise between under- and over-predicting the transiting planet occurrence, a value of $C = 0.95$ (i.e.\ a $2\upsigma$ confidence interval) is adopted for all rates and uncertainties presented and discussed hereafter. Because the 20\,s-cadence data have a higher detection sensitivity than the 120\,s data, the optimum effective sample size is a combination of data from both cadences. The combined uncertainty can be computed via error propagation, and both of these steps were executed following eqs.~8 and 9 in \cite{vansluijs+vaneylen18-1}.

\subsection{Single transit detection}\label{single_transit_method}

The non-periodic nature of single transits requires an alternative detection strategy. To search for such signals in the {\em TESS} light curves, the data were combed for outlier flux points, using a sliding filter of size 8\,h (a factor of five greater than the maximum expected duration for a Jupiter-size planet with a 200\,d period). Fluxes more than four standard deviations below the filter median were flagged as candidate single transits. This threshold represents a good balance between sensitivity to shallow transits and a manageable number of spurious detections. These were subsequently filtered by only accepting correlated changes in flux, with a minimum number of four and 12 data points in the event for 120 and 20\,s cadence light curves, respectively. This minimum-case signal was chosen to be representative of a transiting Mars-size planet on a 45\,d orbit. 

\begin{figure}
\includegraphics[width=\columnwidth]{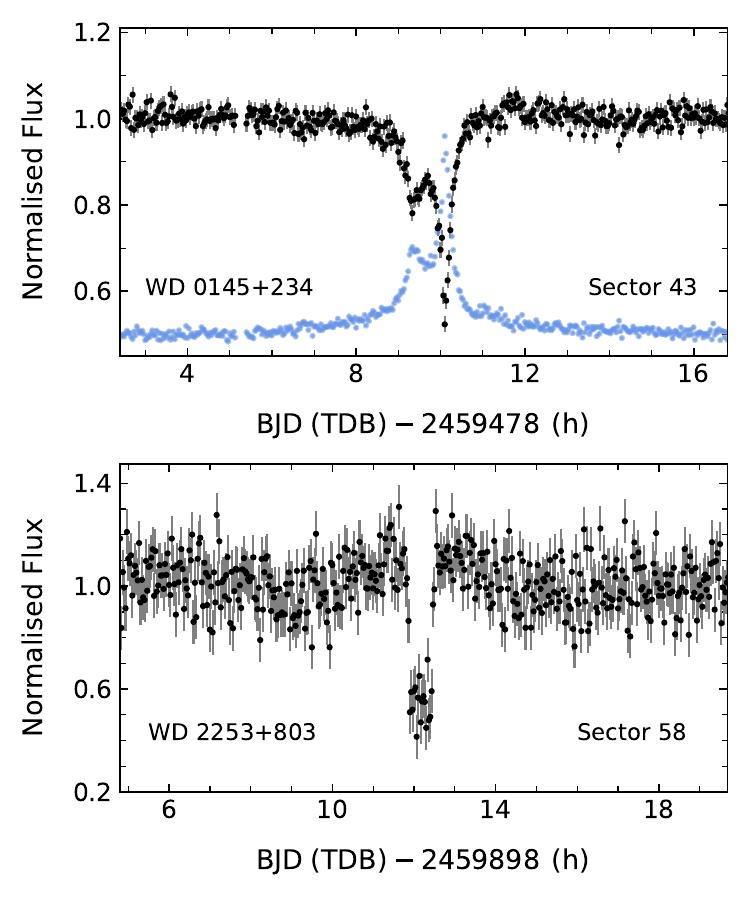}
\caption{Portions of the two {\sc pdcsap} {\em TESS} light curves flagged in the single transit search, where the relevant sector in which the dimming event occurs is displayed for each star. The correctly-scaled background flux for WD\,0145+234 is overplotted in blue, mirroring the over-subtraction resulting from a passing asteroid. The cause of the apparent dimming for WD\,2253+803 is from a pointing excursion, and results in similar features -- some dimming but others brightening -- occurring at the same time in different pixels on the array.}
\label{single_transit_LC}
\end{figure}

Adjacent arrays were then compared to reduce the possibility of long-term systematics or correlated noise being flagged as candidate events. This detection routine was tested on a sample of 200 {\em TESS} light curves with injected single transit signals, and produced a 90~per cent true positive rate with zero false negatives for (synthetic) transit depths larger than 4~per cent. The single transit detection algorithm was then run for the ground-based data and {\em TESS} light curves. No single transits, consistent with a planetary body, were confidently detected in either the ground-based or {\em TESS} data. However, two false positives, WD\,0145+234 and WD\,2253+803, were flagged in this search, and are shown in Fig.~\ref{single_transit_LC}. The double-dip feature present in the Sector 43 light curve of WD\,0145+234 is caused by a spurious increase in the sky estimate in the {\sc pdcsap} pipeline aperture photometry, which is likely caused by a passing asteroid. As is the case with WD\,2230--125 (see Fig.~\ref{ucam_uspec_TESS_LCs}), the estimated speed of the passing objects are consistent with the orbits of asteroids, which are slower than the earth (25\,km/s) owing to wider orbits (although the relative speed to the {\em TESS} satellite can be slightly smaller or larger). The speeds of these objects are not consistent with satellites. The single light curve dip apparent for WD\,2253+803 in Sector 58 was caused by an instrumental pointing excursion lasting approximately 30\,min, during which fluxes were re-distributed into adjacent pixels, resulting in corresponding dimming and brightening events in different parts of the array. Although these features are not true dimming events in the target star light curves, their detection underscores the ability of this method to recover single transits.

\begin{figure}
\includegraphics[width=\columnwidth]{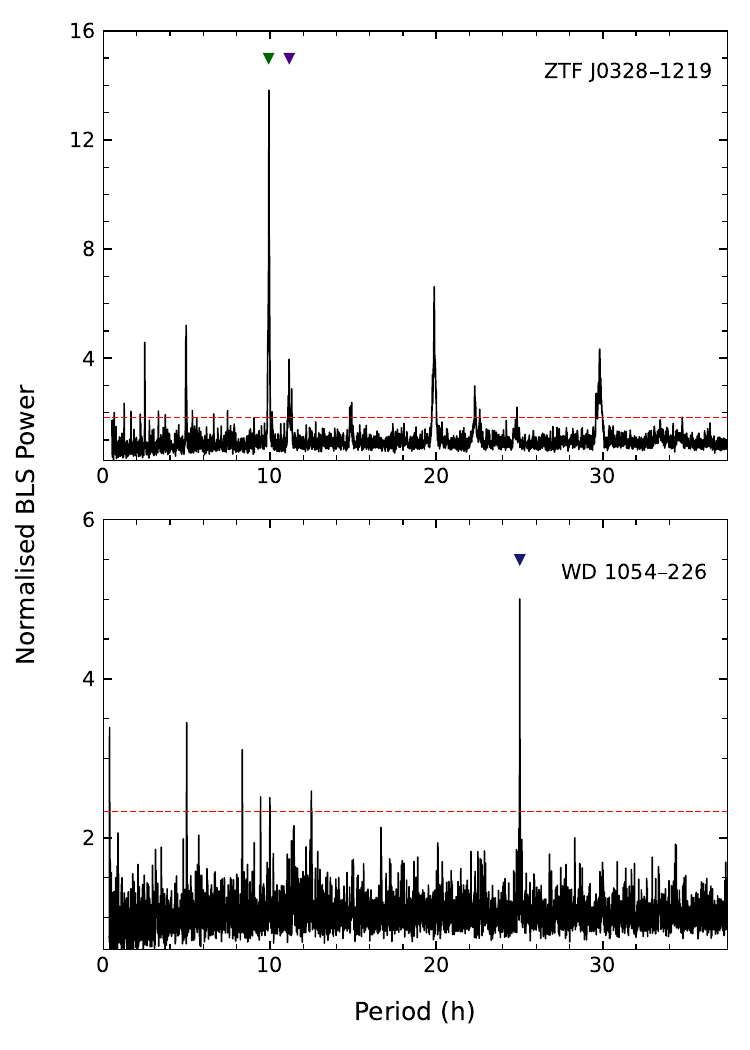}
\caption{The normalized BLS periodograms of all available sectors of the 120\,s-cadence light curves for ZTF\,J0328--1219 and WD\,1054--226. Each panel marks the fundamental signals using colored triangles, where a number of peaks corresponding to true harmonics can be seen at shorter period values (peaks at integer multiples of the fundamental periods are artefacts resulting from the BLS algorithm). The periodograms have been normalized by dividing the BLS power spectra by a $2^{\rm nd}$-order polynomial fit to each respective spectrum. The red dashed horizontal lines indicate the normalized, 0.1\,per cent probability, false alarm threshold for each periodogram.}
\label{bls_wd1054_ztfj0328}
\end{figure}

\section{Results and Discussion}\label{s4}

This section outlines the results of the transit search, and derives upper limits for transiting planet occurrence based on their null detection. The {\em TESS} results are compared with related studies and simple predictions based on transit probability, applied to polluted white dwarfs.

\subsection{Recovered transiting and variable systems}

The search blindly recovered two of three targets with known irregular transits from debris clouds via their {\em TESS} light curves; ZTF\,J0328--1219 \citep{vanderboschetal21-1} and WD\,1054--226 \citep{farihietal22-1}, but not WD\,1145+017 \citep{vanderburgetal15-1}.  Normalized BLS periodograms computed for the {\sc pdcsap} light curves of both stars with detected transits are shown in Fig.~\ref{bls_wd1054_ztfj0328}. Of the remaining, metal-rich white dwarfs with known transiting debris, SDSS\,J0107+2107, ZTF\,J0139+5245 and SBSS\,1232+563 are too faint to be included in the TIC and are not part of the target sample. WD\,1856+534, which has regular transits from a candidate giant planet \citep{vanderburgetal20-2}, does not meet the sample selection criteria because it is not metal-polluted. Thus, a nominal fraction of $0.8_{-0.4}^{+0.6}$~per cent of polluted white dwarfs exhibit detectable transits in {\em TESS}, where the median value and limits (16$^{\rm th}$ and 84$^{\rm th}$ percentiles) are derived using a beta distribution with a flat prior and a binomial likelihood.

\begin{figure}
\includegraphics[width=\columnwidth]{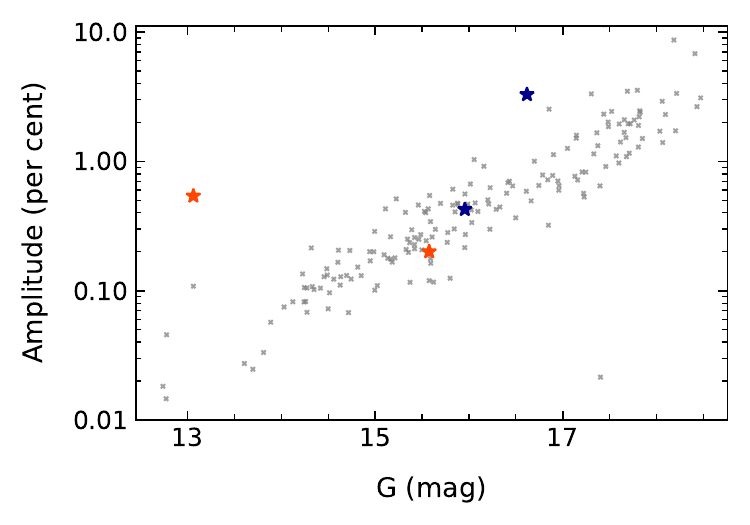}
\caption{The 0.1~per cent probability, false-alarm thresholds for the metal-polluted stars in the {\em TESS} 120\,s sample with {\sc crowdsap} $>0.5$ (grey crosses), and the variability amplitudes for the recovered transiting systems (blue stars; ZTF\,J0328--1219 and WD\,1054--226) and pulsators (orange stars; G29-38 and PG\,1541+651). The plot illustrates a general trend of decreasing sensitivity with decreasing source brightness, but nevertheless yields excellent sensitivity to variability for the bulk of objects.}
\label{amplitude_thresholds}
\end{figure}

The non-detection of WD\,1145+017 is notable, as it has been observed to exhibit transit depths of 40 to 60~per cent in observations spanning at least a few years \citep{gaensickeetal16-1,garyetal17-1, izquierdoetal18-1}.  Despite having $T=17.3$\,mag, it has an excellent {\sc crowdsap} of 0.93, hence variations of several per cent should be detectable in {\em TESS} data (\citealt{rickeretal15-1}; see Fig.~\ref{amplitude_thresholds}). Nevertheless, neither BLS nor Lomb-Scargle reveal any significant periodicities, and the null detection suggests a change in the circumstellar activity of the prototype transiting white dwarf. This is corroborated by recently published, ground-based observations that show the transits are greatly diminished in depth and frequency in 2022 and 2023, and notably correlated with long-term, flux-calibrated photometry \citep{aungwerojwitetal24-1}.  Had WD\,1145+017 been detected, the transiting fraction would have been $1.2^{+0.7}_{-0.5}$~per cent, and thus not substantially changed.

The Lomb-Scargle search recovered two known ZZ~Ceti stars that passed the {\sc tess-localize} validation process: ZZ~Psc (G29-38; \citealt{mcgraw75-1}) and KX~Dra (PG\,1541+651; \citealt{vauclairetal00-1}). Notably, the BLS algorithm did not detect either of these pulsators, where the light curve variations are increases from the baseline flux.  Fig.~\ref{amplitude_thresholds} plots the detected amplitudes of the variable polluted stars (both transiting and pulsating), together with the detection thresholds for non-variable {\sc pdcsap} light curves, each computed using Lomb-Scargle. Only stars with {\sc crowdsap} $>0.5$ are included in the plot, because photometric contamination from nearby field stars can superficially augment the variability threshold.  For the 180 stars plotted, the figure demonstrates excellent sensitivity to variable signals with amplitudes smaller than 1~per cent in most cases.

\subsection{Results of the injection and recovery}\label{injecton_results}

\begin{figure*}
\includegraphics[width=\textwidth]{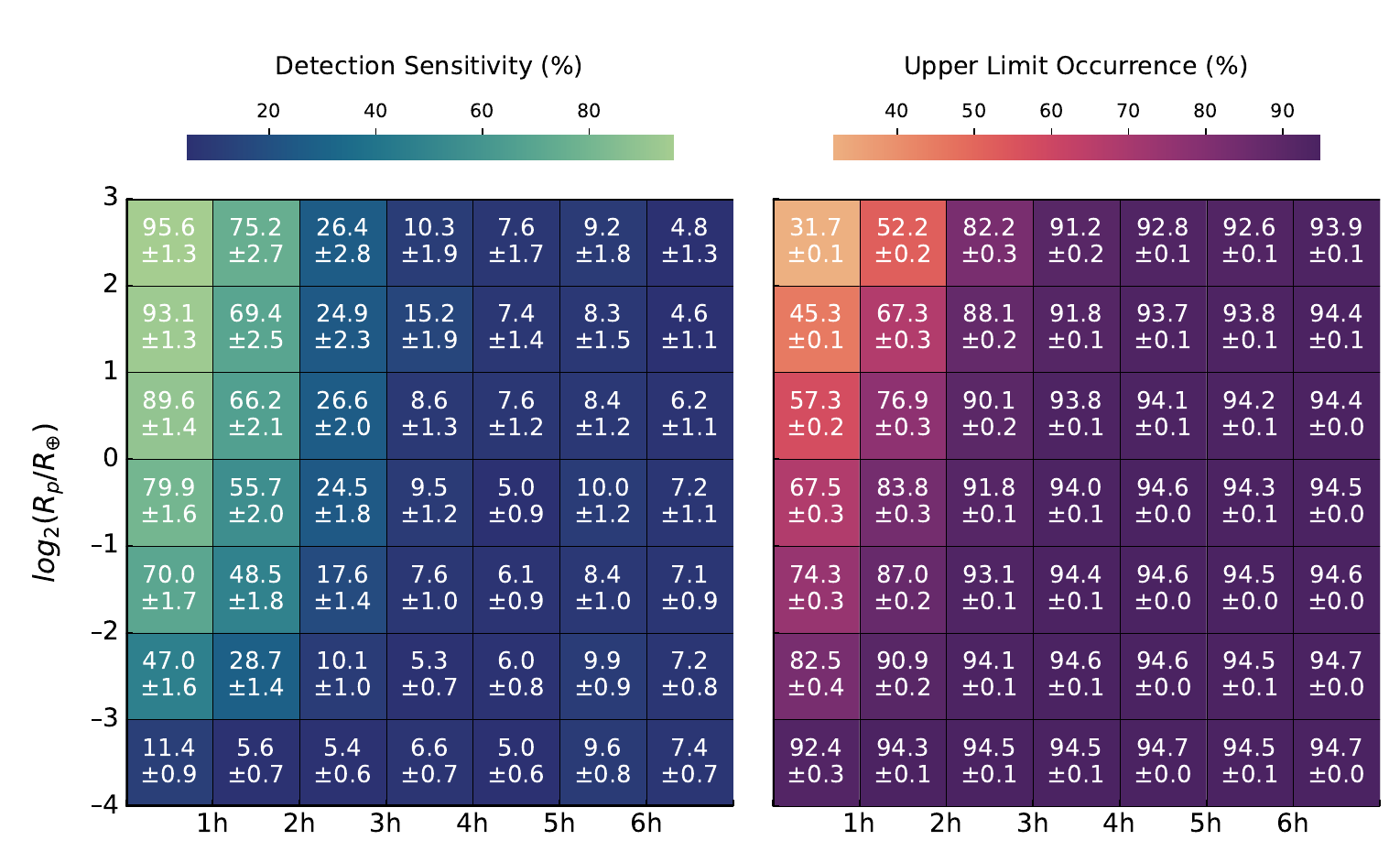}
\includegraphics[width=\textwidth]{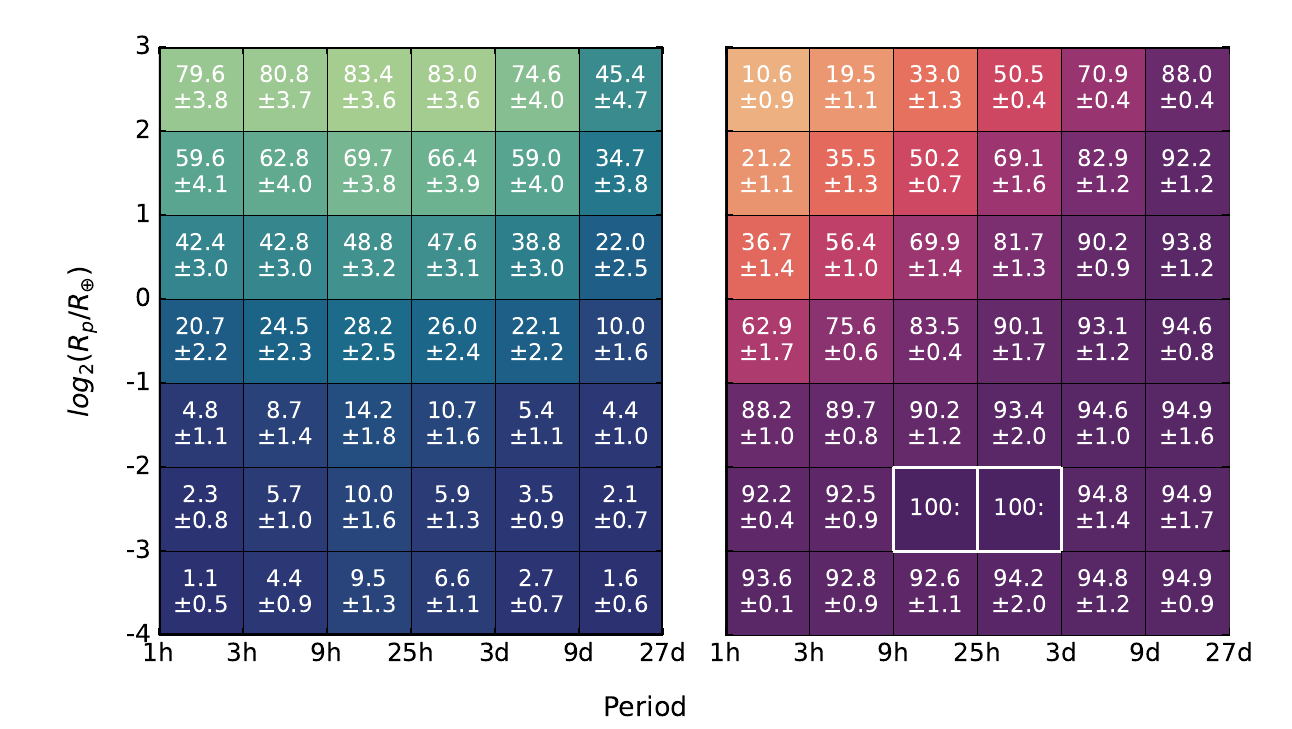}
\caption{{\em Left panels}: The detection sensitivity for the ground-based (top) and {\em TESS} observations (bottom), as a function of planet radius and orbital period, where both sets of simulations cover the same range of planetary radii. The {\em TESS} simulations constrain a wider range of orbital periods, where the detection sensitivity was calculated based on the weighted average of the 120 and 20\,s data sets. {\em Right panels}: The corresponding, upper limit frequencies for transiting planets around polluted white dwarfs. The upper limit occurrences for the two {\em TESS} cadences were combined following the methodology outlined in \citet{vansluijs+vaneylen18-1}, and the uncertainties computed using Eq.~\ref{sigma_max_occurrence_null}. The two bins outlined in white each contain a transit detection, and are consistent with $n/N'=100$~per cent occurrence, but only weakly constrained (see Section~\ref{geometry}). For orbital periods within a few hours and thus interior to the Roche limit, the results apply to debris clouds with an effective radius defined by the simulated transit depths.}
\label{all_results}
\end{figure*}

The resulting detection sensitivities, transit occurrences and upper limits for all samples are displayed in Fig.~\ref{all_results}. The errors for the detection sensitivities were calculated using bootstrap re-sampling (this procedure is outlined in Section \ref{transit_search}). The results of the injection and recoveries were re-sampled a total of 10\,000 times per period-radius bin, and the standard deviation of the re-sampled distribution encodes the error that arises owing to a finite number of injections.  The ground- and space-based results have some common trends: for synthetic planets with larger radii and shorter orbital periods, the transit depths are larger and there are more transits in a given light curve, thus enabling higher detection sensitivity, while the opposite is true for simulated planets with smaller radii and longer periods. Because transit probability is also higher for closer orbits, Eq.~\ref{max_occurence_null} provides stronger constraints for such planets. The shorter light curve duration (median 4.1\,h) and sample size for the ground-based observations result in weak upper limits for transits with periods longer than around 2\,h.  This effect can be seen by comparing the strongest occurrence constraints for the ground-based data (31.7~per cent for $P<1$\,h) with a similar period-radius bin for the {\em TESS} sample (10.6~per cent for 1\,h $< P < 3$\,h). The visual inspection process was not applied to the synthetic transits, hence there may be potential imperfections in the recovery, and thus detection sensitivity. Because self-gravitating objects such as planets and brown dwarfs cannot have orbits interior to the Roche limit, the ground-based results presented in Fig.~\ref{all_results} are better suited to describe the occurrence of circumstellar debris around polluted white dwarfs. However, for orbits longer than several hours, the space-based results are applicable to the occurrence of debris clouds and self-gravitating planetary bodies.

\subsection{Comparison with prior work}

The first effort to constrain the occurrence of transiting planets around white dwarfs used data from the Wide-Angle Search for Planets survey (WASP) where the analysis consisted of 194 white dwarfs with 12\,mag $<V<15$\,mag that were selected without regard to spectral type, mass, or multiplicity \citep{faedietal11-1}.  Assuming the sample are all products of single star evolution, the WASP data most strongly constrain the frequency of brown dwarf and giant planetary companions to no more than 10~per cent for periods less than 0.2\,d. The results presented here for {\em TESS} provide similar, focused limits for polluted white dwarfs in corresponding period-radius bins, but also provide modest constraints for planets at greater distances, and significantly smaller radii.

A search for planetary transits around white dwarfs was also performed using data from the Panoramic Survey Telescope and Rapid Response System (Pan-STARRS; \citealt{fultonetal14-1}). The size of the white dwarf sample in this study is significantly uncertain, with 661 strong candidates identified by reduced proper motion, but the bulk of the 1718 candidates selected using color-color cuts that likely include contaminants.  Nevertheless, the study found no transit candidates for light curves with a few 10$^3$ epochs per star, with four to eight consecutive 240\,s exposures per night.  The study reports an upper limit of 0.4~per cent for transiting planets with $R_{\rm p} \gtrsim 2\,{\rm R}_{\oplus}$ and semimajor axes smaller than 0.02\,au. Keeping in mind that this upper limit frequency may be significantly underestimated based on the uncertain sample size of bona fide, single white dwarfs, {\em TESS} compares more favorably in the bulk of the parameter space in Fig.~\ref{all_results}, especially for planets smaller than Earth at any orbital distance.

A search for transiting substellar objects around 1148 high-probability white dwarfs in {\em K2} recovered nine eclipsing binaries, and the transits towards WD\,1145+017 \citep{vansluijs+vaneylen18-1}. An upper limit frequency of 82~per cent was placed for planets larger $R_{\rm p}\,\approx\,0.5\,{\rm R}_{\oplus}$ with orbital periods less than a few days, and is comparable to 84~per cent found with the {\em TESS} sample of single, polluted white dwarfs.  However, {\em K2} is more sensitive to planets with $P < 15.6$\,h and $R_{\rm p} > {\rm R}_{\oplus}$), assuming a significant fraction of the sample white dwarfs are intrinsically single. With the crude assumption that the values reported in each {\em K2} period-radius bin are representative of the entire bin range, then orbits between 10 and 27\,d are roughly equally-well constrained by {\em TESS}, with weak constraints on smaller planetary radii than {\em K2}.

Using both forward modelling and hierarchical Bayesian inference, it has recently been predicted that transiting rocky planets around white dwarfs are exceedingly rare \citep{kipping24-1}. While consistent with the results presented here for transits of self-gravitating bodies, the prediction does not seem to account for engulfment, which is primarily responsible for their absence.  In fact, direct engulfment is expected to clear all planetary bodies out to several au following both the first-ascent and asymptotic giant branches \citep{mustill+villaver12-1, nordhaus+spiegel13-1}. Thus, the candidate transiting planet WD\,1856+534b is indeed a surprise, but likely accounted for by Lidov-Kozai migration \citep{munozetal20-1,oconnoretal21-1}.  Furthermore, the prediction does not capture the necessarily frequent orbital injection of rocky planetesimals that are required for the observed debris discs and white dwarf photospheric pollution. {\em This planetary material must have an origin further out to avoid engulfment during prior stellar evolution}.  If the two stars with irregular transits recovered in this work are indicative of a 100~per cent occurrence of closely-orbiting planetary material, then a similar fraction of polluted white dwarfs are host to distant planets that are dynamically responsible for the star-grazing orbits required to approach the Roche limit.

\subsection{Debris transits consistent with frequent occurrence}\label{geometry}

Because all white dwarfs in this study are metal-polluted, the accretion of planetary material is either ongoing or recent, and thus most or all targets are expected to have circumstellar debris, where two stars were detected to have associated transits. If all polluted white dwarfs have debris discs orbiting near their Roche limits ($1.2\,R_\odot$), and these are sufficiently opaque and vertically extended to marginally result in detectable transits ($p_{\rm transit}=1.4$~per cent), then 4.4 events are expected in a sample of 313.  These predictions are within the error bars of the 0.8$^{+0.6}_{-0.4}$~per cent transit fraction and 2.0 events recovered in this study.  The results are broadly compatible with a model where all polluted white dwarfs have debris orbiting near the Roche limit, and thus their detections are the result of favourable viewing geometry.

The bulk of occurrences presented in Fig.~\ref{all_results} are upper limits relevant to null detections, with the exception of the two bins where transiting debris is identified (outlined in white). Owing to the modest detection sensitivity and small transit probability, both bins with detected transits have $n/N'$ that saturates at 100~per cent. Alternatively, using a modified Eq.~2 to compute the upper limit occurrence given a single detection also yields unity. Despite a small effective sample size at these radii and periods, these occurrence values are consistent with circumstellar material orbiting all polluted white dwarfs, and in agreement with the prediction of transit probability above.

\subsection{Constraints on exocomets}

There are several models that have considered Oort-like comet clouds to account for metal pollution in white dwarfs, where the first preceded the discovery of their debris discs \citep{alcock86-1}.  All studies agree that a modest fraction of distant comets should remain bound, even in the case of asymmetric mass loss during stellar evolution \citep[e.g.][]{verasetal14-4}. Given the potential numbers, exocomets might significantly contribute to metal pollution, where simulations demonstrate they can be perturbed from several thousand au into the Roche limit of a white dwarf, either by a natal kick \citep{stoneetal15-1}, or from Galactic tides \citep{oconnoretal23-2, pham+rein24-1}. Extrasolar comets are apparent toward young stars like $\upbeta$~Pic via time-series spectroscopy \citep{ferletetal87-1}, and more recently via transits in that system \citep{ziebaetal19-1}, and others \citep{rappaportetal18-2}.  It thus seems plausible and worthwhile to consider their survival and possible transits near periastron around polluted white dwarfs, particularly because cometary comae are typically Earth-sized or larger, and so their transits should be straightforward to detect in white dwarf light curves.  

The probability that a transit is detected within a sample of $N=313$ white dwarfs is 

\begin{equation}
    p_{\rm detect} = p_{\rm transit} \times N \times m \times f_{\rm orbit} \times f_{\rm coma},
\end{equation}

\noindent
where $m$ is the number of exocomets hosted by each star, $f_{\rm orbit}$ is the fraction of the orbital period observed (e.g.\ by {\em TESS}), and $f_{\rm coma}$ is the fraction of the orbit over which a coma (i.e.\ outgassing) is present and detectable. To estimate this overall probability, the following parameters are adopted: a semimajor axis of 3000\,au \citep{duncanetal87-1}, periastron at $1\,R_\odot$, $p_{\rm transit} = (R_* + R_\oplus)/a/(1-e^2)$ for Earth-sized comae, $f_{\rm orbit} = 8.5 \times 10^{-7}$ for the median observing coverage of 2.3 {\em TESS} sectors in this study, and $f_{\rm coma} = 5.5 \times 10^{-8}$ for activity that begins around 150\,K (i.e.\ interior to 0.15\,au) for a white dwarf luminosity of $2\times10^{-3}\,L_\odot$. Under the assumption that each exocomet orbit approaches the Roche limit, $m = 2 \times 10^{12}$ comets are needed to achieve $p_{\rm detect}=0.5$, and thus a reasonable chance of detecting one in transit.  This ad hoc scenario would require all polluted white dwarfs to have bountiful Oort-like clouds, each dominated by star-grazing orbits. While unrealistic, this calculation nevertheless provides a detection benchmark, where it is notable that a recent model predicts distant planetesimal belts can be excited to high eccentricities following white dwarf natal kicks \citep{akibaetal24-1}.

There are models that specifically calculate the rate at which Oort-cloud analogs are delivered to the Roche limit of white dwarfs, including the effects of Galactic tides, as well as the mitigating influences of companions and stellar evolution \citep{oconnoretal23-2,pham+rein24-1}.  The most optimistic of these predict that comets may reach the innermost regions at rates as high as $\sim2\times10^{4}$\,Myr$^{-1}$, in the absence of a companion and without substantial loss of comets during post-main sequence evolution (each of which can reduce this number by an order of magnitude).  During the median observing coverage of 2.3 {\em TESS} sectors for the 313 white dwarfs in the sample, this rate predicts each star could have up to 0.0035 comets on such steeply infalling orbits, where a fraction of 0.011 of these would transit.  Overall, there would have been a 1~per cent chance at best to detect an exocomet transit in the space-based sample, and this probability is likely less favourable based on the aforementioned model considerations. Thus, the lack of a cometary transit here does not significantly constrain the prevalence of extrasolar Oort clouds, but future samples that are orders of magnitude larger should be more informative.

\section{Conclusion and outlook}

This study searched for transiting planetary systems around polluted white dwarfs using ground-based observations with ULTRACAM and ULTRASPEC, and space-based data from {\em TESS}. Both the BLS and Lomb-Scargle algorithms were used to search for periodic signals in the target light curves, and two systems known to have periodic dimming events associated with debris discs were blindly recovered. The calculated detection sensitivities provide upper limits for the occurrence of planets and planetesimals ranging from dwarf to Kronian planetary radii.  The nominal fraction of polluted white dwarfs with detectable transits is $0.8_{-0.4}^{+0.6}$~per cent.

The 40\,pc sample of white dwarfs contains 1076 stars with 99.3~per cent spectroscopic completeness \citep{obrienetal24-1}. Assuming a pollution fraction of 30~per cent \citep{zuckermanetal10-1, koesteretal14-1}, and the transit incidence found here, then $2.6_{-1.6}^{+2.0}$ metal-rich white dwarfs within this volume are expected to exhibit transits. Currently, WD\,1054--226 is the only polluted white dwarf with transits within 40\,pc (ZTF\,J0328--1219 is just outside this volume), and it is therefore possible that one or two additional nearby, polluted white dwarfs will have transiting debris discs. 

While models for white dwarf pollution require major planets to dynamically perturb asteroids to the Roche limit, these are expected to have semimajor axes of several to tens of au \citep{bonsoretal11-1, frewen+hansen14-1}.  However, planets might be born of recycled disc material, as a result of viscous spreading beyond the Roche limit for sufficiently massive debris discs \citep{vanlieshout18-1}, or achieve temporary, star-grazing orbits due to unpacking of previously stable, multi-planet systems \citep{veras+gaensicke15-1}. While none are detected here, the planetary constraints are modest outside of orbital periods of a few days.

There is a basic necessity for planets in polluted white dwarf systems, as the tidally-disrupted planetesimals responsible for debris discs and pollution require dynamical perturbations.  Moreover, these orbits have to be sufficiently distant to escape engulfment during stellar evolution, and will likely have expanded further owing to stellar mass loss \citep{jeans24-1}.  Such widely-bound planets may be directly imaged by {\em JWST} if they are at least as massive as Jupiter at favorable angular separations \citep{carteretal23-1}.  However, the requisite number of observations to achieve a detection will be challenging. 

It is likely that the most promising avenue for discovering planets orbiting (polluted) white dwarfs will be astrometric detection with {\em Gaia}. The fraction of white dwarfs that host planets detectable with {\em Gaia} astrometry is predicted to be as high as 0.1~per cent \citep{perrymanetal14-1, sandersonetal22-1}, yielding an order of ten planets on au-scale orbits. There are currently three white dwarfs with candidate planets that have been inferred through a combination of {\em Gaia} and {\em Hipparcos} astrometry as proper motion anomalies \citep{kervellaetal22-1}. Furthermore, {\em Gaia} DR4 should provide epoch photometry for millions of stars, and increase the number of white dwarfs with space-based light curves by over two orders of magnitude compared to {\em TESS}.  While the cadence and photometric precision of {\em Gaia} will remain inferior to {\em TESS}, the sheer numbers should result in new transiting systems.  Rubin-LSST will obtain light curves for up to 10$^8$ white dwarfs, but the observing cadence and baseline may complicate immediate characterization of any transits.  Nevertheless, if the 0.8~per cent fraction determined here is correct to an order of magnitude, then several thousands of transiting systems await detection. Even if only a tiny fraction of these are sufficiently bright for follow up, this can transform the field from the study of individual stars to statistical analyses of a population.

\section*{Acknowledgements}

The authors thank E.~M.~Bryant for useful discussions about injection and recovery simulations.  A.~Robert acknowledges support from a Science and Technology Facilities Council studentship. This work was supported by Naresuan University (NU), and National Science, Research and Innovation Fund (NSRF) (grant No.\ R2566B021). This project has received funding from the European Research Council (ERC) under the European Union’s Horizon 2020 research and innovation programme (grant agreement No.\ 101020057).The authors acknowledge the European Southern Observatory for the award of telescope time via programs 0100.C-0546, 0102.C-0700, 0103.C-0247, and 105.209J. This work has made use of data obtained at the Thai National Observatory on Doi Inthanon, operated by NARIT.  ULTRACAM and ULTRASPEC are funded by the Science and Technology Facilities Council (grant ST/V000853/1). This paper utilizes data collected by the {\em TESS} mission, which is funded by the NASA Explorer Program, including targets from Guest Investigator programs: G011113, G011179, G011203, G011288, G011294, G022017, G022057, G022065, G022077, G022158, G022176, G022217, G03046, G03124, G03126, G03207, G03221, G03252, G03274, G04036, G04056, G04091, G04106, G04123, G04137, G04171, G04177, G04200, G04209, G04211, G04219, G04236, G04240, G05003, G05057, G05071, G05081, G05084, G05115, G05148.

\section*{Data availability}
ULTRACAM and ULTRASPEC data are available on reasonable request to the authors. {\em TESS} data are available through the Mikulski Archive for Space Telescopes. 



\bibliographystyle{mnras}
\bibliography{references}





\appendix

\section{Ground-based observation summary}

\begin{table}
\caption{Ground-based target observations.}\label{tab:obs}
\begin{tabular}{ccccc}
\hline
WD 			&{\em Gaia} DR3 	         &$G$ 		&Light curve &Duration\\
			&					         &(mag)		&S/N		 &(h)\\

\hline
0047+190 	&2788992130973584640 		&16.0 		&176 		&2.6\\
0106--328 	&5026963661794939520 		&15.4 		&191 		&2.8\\
0110--565 	&4913589203924379776 		&15.8 		&218 		&2.6\\
0145+234 	&291057843317534464 		&14.0 		&82 		&5.0\\
 			&2489275328645218560 		&16.4 		&109 		&4.0\\
0300--013 	&5187830356195791488 		&15.5 		&38 		&5.1\\
0307+077 	&13611477211053824 		    &16.2 		&78 		&6.0\\
0322--019 	&3261591090771914240 		&15.9 		&186 		&2.5\\
 			&53278867446391040 		    &16.6 		&97 		&2.0\\
0408--041 	&3251748915515143296 		&15.6 		&53 		&6.0\\
0420--731 	&4653404070862114176 		&15.5 		&221 		&2.6\\
0435+410 	&203931163247581184 		&14.8 		&73 		&2.4\\
0446--255 	&4881758307940843008 		&16.9 		&98 		&2.7\\
0449--259 	&4881672923991046400 		&16.4 		&144 		&2.7\\
 			&4823042355496728832 		&17.5 		&10 		&1.3\\
0536--479 	&4795556287084999552 		&15.6 		&262 		&2.8\\
0735+187 	&671450448046315520 		&17.7 		&16 		&6.4\\
0738--172 	&5717278911884258176 		&13.0 		&305 		&3.0\\
0843+516 	&1029081452683108480 		&16.1 		&22 		&10.7\\
1041+092 	&3869060540584643328 		&17.3 		&30 		&4.9\\
1054--226 	&3549471753507182592 		&16.0 		&69 		&22.6\\
1202--232 	&3489719481290397696 		&12.7 		&301 		&3.0\\
1225--079 	&3583181371265430656 		&14.7 		&34 		&4.6\\
1226+110 	&3904415787947492096 		&16.4 		&17 		&2.2\\
1349--230 	&6287259310145684608 		&16.5 		&194 		&2.0\\
1350--162 	&6295510766956001024 		&17.0 		&74 		&2.0\\
1455+298 	&1281989124439286912 		&15.5 		&63 		&4.2\\
1532+129 	&1193520666521113344 		&15.6 		&280 		&1.8\\
1554+094 	&4454599238843776128 		&18.6 		&48 		&2.5\\
1620--391 	&6018034958869558912 		&11.0 		&79 		&5.5\\
2138--332 	&6592315723192176896 		&14.4 		&174 		&2.0\\
2230--125 	&2601689466188247808 		&16.1 		&233 		&2.6\\
2354+159 	&2773136383028850688 		&15.7 		&182 		&2.5\\
\hline
\end{tabular}
{\em Note.}  The fourth column is the light curve mean divided by the standard deviation after de-trending.
\end{table}

\bsp	
\label{lastpage}
\end{document}